\documentclass[a4paper, amsfonts, amssymb, amsmath, reprint, showkeys, nofootinbib, twoside, superscriptaddress]{revtex4-2}
\usepackage[english]{babel}
\usepackage[utf8]{inputenc}
\usepackage[colorinlistoftodos, color=green!40, prependcaption]{todonotes}
\usepackage{amsthm}
\usepackage{mathtools}
\usepackage{physics}
\usepackage{xcolor}
\usepackage{graphicx}
\usepackage{subcaption}
\usepackage[left=18mm,right=18mm,top=35mm,columnsep=15pt]{geometry} 
\usepackage{adjustbox}
\usepackage{placeins}
\usepackage[T1]{fontenc}
\usepackage{lipsum}
\usepackage{csquotes}

\usepackage[pdftex, pdftitle={Article}, pdfauthor={Author}, hidelinks]{hyperref} 

\usepackage[justification=raggedright,format=plain]{caption}

\usepackage[nolist,nohyperlinks]{acronym} 

\usepackage{color}
\usepackage[T1]{fontenc}

\newcommand{\QETLABS}{Quantum Engineering Technology Labs, H. H. Wills Physics Laboratory \& Department of Electrical and Electronic Engineering, University of Bristol, Tyndall Avenue, Bristol, BS8 1FD UK}
\newcommand{\QECDT}{Quantum Engineering Centre for Doctoral Training, H. H. Wills Physics Laboratory \& Department of Electrical and Electronic Engineering, University of Bristol, Tyndall Avenue, Bristol, BS8 1FD, UK}
\newcommand{\KETS}{KETS Quantum Security, Station Road Workshops, Bristol, BS15 4PJ, UK}
\newcommand{\RCQI}{RCQI, Institute of Physics, Slovak Academy of Sciences, Dúbravská Cesta 9, 84511, Bratislava, SK}

\begin{document}
\title{Physical security of chip-based quantum key distribution devices}

\author{Friederike Jöhlinger}
    \email[]{f.joehlinger@outlook.com}
    \affiliation{\QETLABS}
    \affiliation{\QECDT}

\author{Henry Semenenko}
    \affiliation{\QETLABS}
    \affiliation{\QECDT}
    
\author{Philip Sibson}
    \affiliation{\QETLABS}
    \affiliation{\KETS}

\author{Djeylan Aktas}
    \affiliation{\QETLABS}
    \affiliation{\RCQI}

\author{John Rarity}
    \affiliation{\QETLABS}

\author{Chris Erven}
    \affiliation{\QETLABS}
    \affiliation{\KETS}

\author{Siddarth Joshi}
    \email[]{sk.joshi@bristol.ac.uk}
    \affiliation{\QETLABS}

\author{Imad Faruque}
    \email[]{imad.faruque@bristol.ac.uk}
    \affiliation{\QETLABS}

\date{\today} 

\begin{abstract}

The security proofs of the \ac{QKD} protocols make certain assumptions about the operations of physical systems. Thus, appropriate modelling of devices to ensure that their operations are consistent with the models assumed in the security proof is imperative. In this paper, we explore the \ac{THA} using \ac{MDI} \ac{QKD} integrated photonic chips and how to avoid some of the security vulnerabilities using only on-chip components. 
We show that a monitor photodiode paired appropriately with enough optical isolation, given the sensitivity of the photodiode, can detect high power sniffing attacks. We also show that the placement of amplitude modulators with respect to back reflecting components and their switching time can be used to thwart a \ac{THA}.

\end{abstract}

\maketitle

\begin{acronym}
 \acro{AMZI}{asymmetric Mach–Zehnder interferometer}
 \acro{AWG}{arbitrary waveform generator}
 \acro{BS}{beamsplitter}
 \acro{DBR}{distributed Bragg reflector}
 \acro{DOS}{denial-of-service}
 \acro{DFB}{distributed feedback}
 \acro{EOPM}{electro-optic phase modulator}
 \acro{HHI}{Heinrich-Hertz-Institut}
 \acro{InP}{indium phosphide}
 \acro{LDA}{laser damage attack}
 \acro{LSA}{laser seeding attack}
 \acro{MDI}{Measurement Device Independent}
 \acro{MMI}{multi-mode interferometer}
 \acro{MZI}{Mach–Zehnder interferometer}
 \acro{OBR}{optical backscatter reflectometer}
 \acro{PBS}{polarising beam splitter}
 \acro{PDK}{process design kit}
 \acro{PD}{photodiode}
 \acro{PIC}{photonic integrated chip}
 \acro{QKD}{Quantum Key Distribution}
 \acro{SSC}{spot-size converter}
 \acro{TE}{transverse electric}
 \acro{THA}{Trojan horse attack}
 \acro{TM}{transverse magnetic}
 \acro{TOPM}{thermo-optic phase modulator}
\end{acronym}

\section{Introduction} \label{sec:intro}
    
\acf{QKD} is a key exchange protocol that provides provably secure communication that derives from the laws of quantum mechanics and thus is safe even with the advent of the most powerful quantum computer. There are various \ac{QKD} protocols such as BB84, E92, and \ac{MDI} \ac{QKD} \cite{BB84,Lo1999,Braunstein2012,Lo2012} - each requires a security proof with assumptions which can be validated in specific physical implementations thus giving these physical systems advantages for implementing those protocols. \ac{MDI} \ac{QKD} is of particular interest because the protocol is inherently secure against any attack that targets the detectors -- a typical weakness of several other QKD implementations~\cite{Makarov2006,Rogers2007,Wiechers2011,Sauge2011}. Photonics or optical technology is the most dominant platform for such \ac{QKD} implementations, because photons are ideal for communication and sending encryption keys as they interact very little with the environment and thus can carry the quantum information over a long distance with high fidelity. Also, the infrastructure for sending photons over long distances is already in place in the optical fibre network connecting cities and continents, providing the essential physical layer of the internet. 

\ac{QKD} has been implemented in various photonics platforms such as fibre optics, bulk optics and photonic integrated circuits \cite{Korzh2015,Ursin2007,Sibson2017b,Paraiso2021}. However, imperfections in these physical implementations may prevent fulfilling all the security requirements. This makes practical \ac{QKD} systems vulnerable to security threats often referred to as quantum hacking. There are various proof-of-principle physical demonstrations of quantum hacking and their possible countermeasures \cite{Scarani2014, Sun2015, Jain2015, Vakhitov2001}. Chip-scale implementations are appealing for a wider adaptation of \ac{QKD}, due to their potential for mass manufacturability, their small size which would allow handheld "on the move" \ac{QKD} systems \cite{Lowndes2021}, their low weight and low power transmitter systems which can be reconfigured to implement multiple \ac{QKD} protocols \cite{Sibson2017,Sibson2017b}, and their high-fidelity operations resulting in lower quantum bit error rates \cite{Bunandar2018}. However, from the perspective of quantum hacking, the question remains whether a chip-scale implementation of \ac{QKD} is inherently more or less secure than other photonic platforms.
\begin{figure*}[t!]
    \centering
        \includegraphics[width=\linewidth]{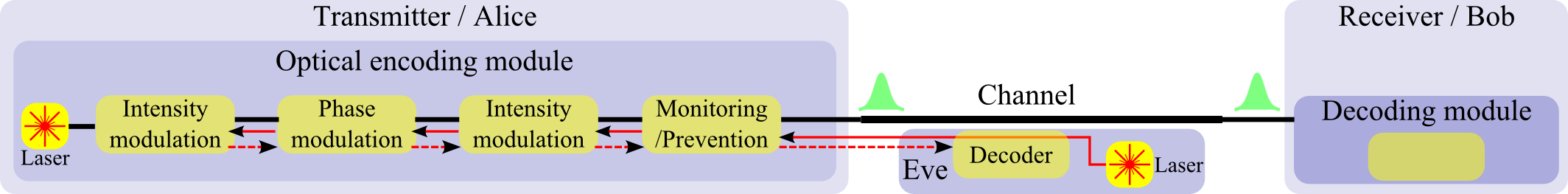}
    \caption{Overview of \ac{THA}. Eve injects light into the transmitter chip which reflects back to Eve, and by measuring the changes in the light, Eve obtains information about the encoding system.}
    \label{fig:overview_THA}
\end{figure*}

There is a great effort in the research community and by commercial companies counteracting security attacks on \ac{QKD} transmitter systems \cite{Sajeed2021, idQ}, for example, the \ac{THA} \cite{Vakhitov2001,Lucamarini2015}, \ac{LDA} \cite{Bugge2014}, and \ac{LSA} \cite{Sun2015}. Most of these attacks require Eve to inject light into the \ac{QKD} transmitter. Therefore, a natural way to counteract these attacks is to protect the outgoing port of the \ac{QKD} transmitter system from any incoming light. This is a standard addition to commercial \ac{QKD} systems, and is usually done using a circulator and photodetector combination for monitoring, or by simply using an optical isolator to prevent Eve from injecting light into the transmitter systems \cite{Jain2014, Jain2015}. Further experiments suggest that using a longer wavelength than the operating wavelength of a commercial \ac{QKD} transmitter, an "invisible" \ac{THA} can be undertaken \cite{Sajeed2017}. Recently, a chip-scale implementation of \ac{MDI} \ac{QKD} in a silicon photonics platform showed that with additional external isolators, a \ac{QKD} transmitter can be made highly secure against the \ac{THA} \cite{Tan2021}. They used multiple isolators in series (with a total of 232 dB isolation), which also provides some protection against a \ac{LDA} attack, however, optical isolators may still be vulnerable to \acp{LDA} at other wavelengths \cite{Makarov2023,Ponosova2022, Ruzhitskaya2021}. Moreover, lower reflections from the chip-scale transmitter and smaller time intervals between the reflection peaks due to the compactness of the components inside the chip make any quantum hacking even harder \cite{Tan2021}.

Circulators and optical isolators are components that are not readily available on-chip, but only as fibre or bulk-optic components. Instead in this paper, we investigate \ac{QKD} transmitters in the \ac{InP} \ac{PIC} platform and devise existing on-chip components as a hacking monitoring and prevention unit. We also analyse the critical parameters of a chip that can reduce the possibility of hacking. We found that the combination of the placement of the on-chip components and the high speed of the \ac{QKD} transmitter actively makes hacking, particularly THA, progressively harder.
  
\section{Trojan Horse Attack on Chips} \label{sec:THA}

We consider the \ac{THA} to demonstrate how Eve could hack a \ac{QKD} transmitter chip. In the \ac{THA}, see for example, in figure \ref{fig:overview_THA}, Eve injects light into an encoding module of the \ac{QKD} system, e.g. the transmitter modules in BB84 or \ac{MDI} \ac{QKD} systems, and measures any back-reflected light. Figure \ref{fig:OBRrefl}(a) shows an \ac{OBR} used to infer information from one of our \ac{InP} \ac{QKD} transmitter chips made by Oclaro, also used in \cite{Sibson2017b,Semenenko2020}. 

The encoding circuit on the chip contains various components from the foundry's building block library. The phase modulation is done by \acp{EOPM}, and the intensity modulation is done using \acp{MZI}, the latter of which are made up of a \ac{MMI} (a type of on-chip beamsplitter) on each end and an \ac{EOPM} in each arm. When Eve's injected light goes through the encoding system it can also receive some encoding, if Eve's timing of the injection of her light is right. When Eve analyses the back-reflected light she can then extract information about the encoding settings used in the \ac{QKD} exchange. This attack does not require Eve to intercept the quantum states sent by Alice, therefore the sender and the receiver (Alice and Bob) will be unaware of this attack. Since Eve is not bound to measurements at single-photon levels, she may obtain full knowledge of the key, and by using the \ac{THA} it is possible for her to completely compromise the security of a \ac{QKD} system.

Here we will show the real risk posed by this attack. We performed initial characterisation of \acp{PIC} using \ac{OBR} to assess the on-chip reflections. We then performed simulations to show how an eavesdropper could compromise a key exchange using the \ac{THA}.

\subsection{OBR measurements}

\begin{figure*}[t!]
    \centering
    \begin{subfigure}[b]{0.45\linewidth}
        \centering
        \includegraphics[width=\linewidth]{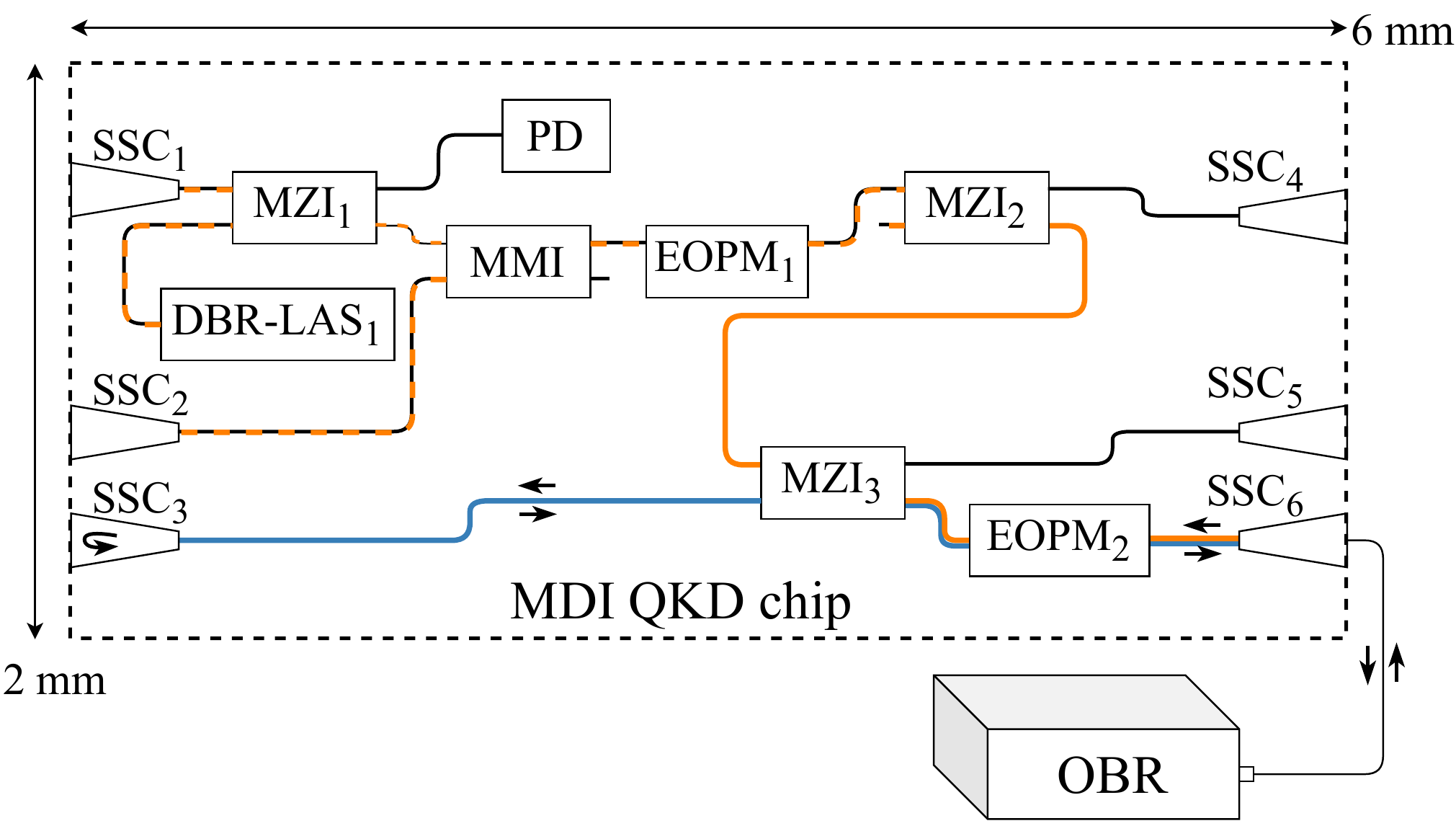}
        \vspace{10pt}
        \caption{}
    \end{subfigure}%
    \begin{subfigure}[b]{0.55\linewidth}
        \includegraphics[width=0.95\linewidth]{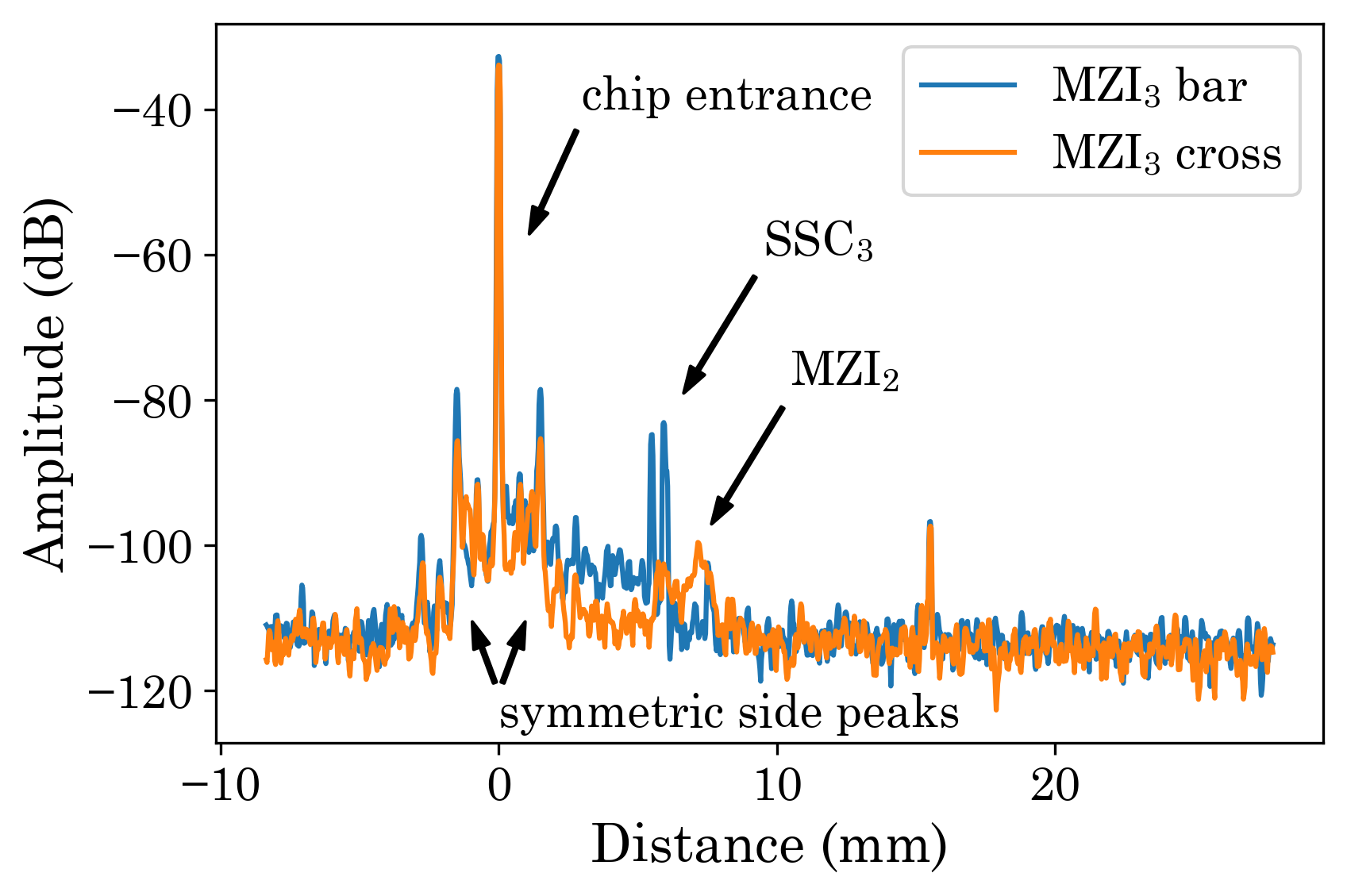}
        \caption{}
    \end{subfigure}
    \caption{\textbf{(a)} Optical circuit of the chip used in the experimental \ac{THA}. The chip contains a circuit consisting of a \ac{DBR} laser, as the \ac{QKD} light source, three \acp{MZI} which can be used as intensity modulators, variable beamsplitters or phase encoders, two stand-alone \acp{EOPM}, one stand-alone \ac{MMI} and one photodiode (PD), for monitoring the source laser power. There are several \acp{SSC} edge couplers connected to this circuit at various points, which can be used for coupling light on or off chip with optical fibres. The blue and orange lines show the paths of the light reflected in the \ac{OBR} measurements. The chip dimensions are $2\times 6$ mm. \textbf{(b)} Two \ac{OBR} traces of the chip, coupled at $\mathrm{SSC_6}$ with a free-space aligned fibre. The main peak at 0 mm corresponds to the entrance to $\mathrm{SSC_6}$, including several symmetric side peaks. When $\mathrm{MZI_3}$ is switched to maximise the output corresponding to the same side (bar configuration `$=$', the blue path) one can see two peaks at $\sim$6 mm from the entrance (reflected from SSC$_{3}$ ). Here, $\mathrm{MZI_3}$ is not letting any information pass from the rest of the circuit to the $\mathrm{SSC_6}$, and hence the large reflections. These peaks are not visible when $\mathrm{MZI_3}$ is switched to maximise output opposite to the input (cross configuration `$\times$', the orange path), instead there are some low reflections around 7.5 mm. Here, $\mathrm{MZI_3}$ is connecting the rest of the circuit to the $\mathrm{SSC_6}$ and we see reflections from other components (e.g. $\mathrm{MZI_2}$) in the encoding circuit.}
    \label{fig:OBRrefl}
\end{figure*}

A schematic layout of the chip used for the \ac{OBR} measurement is shown in figure \ref{fig:OBRrefl}(a). The \ac{QKD} chip can be used to encode time-bin encoded BB84 states, which use two time bins and phase between pulses in the two time bins to encode the 4 states as presented in \cite{Sibson2017}. $\ket{0}$ is encoded as a pulse in only the first time bin, $\ket{1}$ as only a pulse in the latter one. $\ket{+}$ and $\ket{-}$ have pulses in both, with same total intensity over both time-bins together as for $\ket{0}$ and $\ket{1}$. For $\ket{+}$ both pulses are in phase, for $\ket{-}$ the pulses have a $\pi$ phase change between them. These states are created using pulse carving, intensity modulation and phase modulation using the on-chip \acp{EOPM} and \acp{MZI}, which means that measuring the states of each of these components is a potential goal of the \ac{THA}.

Since the \ac{THA} relies on light returning to Eve from the encoding module, it must reflect at some point and the first step is therefore to find any points of reflection inside the encoding module. Figure \ref{fig:OBRrefl}(b) shows reflectometries performed on a \ac{QKD} transmitter chip using a Luna \ac{OBR} 4600. During initial reflectometries, two peaks around 6mm from the entrance were visible. To determine their origin, $\rm MZI_3$ was switched into either maximum throughput in bar ($=$) or in cross ($\times$) direction in next reflectometries. For both reflectometries shown in figure \ref{fig:OBRrefl}(b) light was injected into $\rm SSC_6$, usually the output port of the transmitter for a \ac{QKD} exchange. Here, \acp{SSC} are edge coupling components, which allow coupling between on-chip waveguides and optical fibres. The entrance of the chip is at 0mm, which shows a main peak with several side-peaks, that are mirrored on each side of the main entrance peak, on both \ac{OBR} traces\footnote{In the Luna OBR manual the example traces also include these peaks but their origin is not explained in the manual and thus can be considered as an artefact of the Luna OBR.}.

There is a distinct difference between the two \ac{MZI} settings on the \ac{OBR} traces. When $\rm MZI_3$ was maximised for bar throughput and Eve's light is directed towards $\rm SSC_3$, there are two reflections of about 25dB (above a background of about -113dB) visible at about 6mm into the chip, the distance at which $\rm SSC_3$ is located from the entrance of the chip. These peaks disappear when $\rm MZI_3$ is switched to maximise cross throughput and instead we can see some low and broad peaks at around 7.5 mm into the chip. When considering the direction in which this light goes, it would encounter $\rm MZI_2$ at this distance. There is also a peak around 16 mm after the main entrance peak, however, this peak appears on both traces and is therefore not dependent on the setting of $\rm MZI_3$. The location of the peak suggests that the reflection may not be an artifact on the chip, but  external to it.

Using this reflectometry, we can deduce the setting of $\rm MZI_3$ when used as an intensity modulator using external light. If this \ac{MZI} was used by Alice for e.g. pulse carving for a time-bin encoded scheme, Eve would be able to distinguish between full and empty time bins and learn at least part of the key. This forms the basis of the circuit simulation in the next section.

\subsection{Simulation THA}

\begin{figure*}[t]
    \centering
    \includegraphics[scale=0.65]{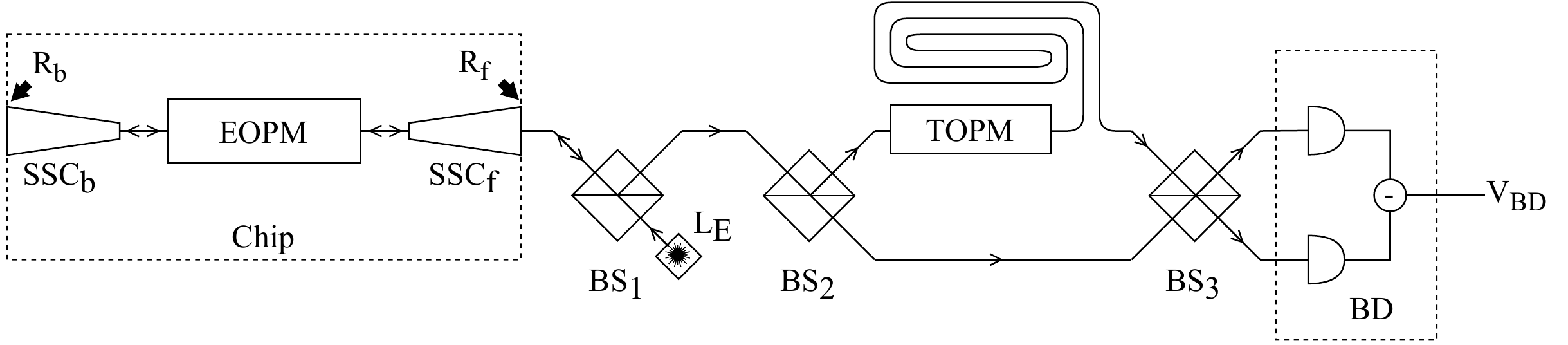}
    \caption[THA on HHI EOPM setup in PICWave simulation]{Simulation setup used for the \ac{THA} on the \ac{HHI} \ac{EOPM}. Alice's chip contains two \acp{SSC}, one at the front ($\mathrm{SSC_f}$) and one at the back ($\mathrm{SSC_b}$) of the chip. In between an \ac{EOPM} was placed. Eve's setup contains three \acp{BS}, for which \ac{MMI} building blocks were used, a \ac{TOPM} in the long arm of the \ac{AMZI} and the balanced detector (BD) building block, whose output $V_{BD}$ is measured. Eve inputs light pulses at 1550 nm into the system using ideal light source $\mathrm{L_E}$. The points of reflection at the front and back of the chip are indicated by $\mathrm{R_f}$ and $\mathrm{R_b}$, respectively.}
    \label{fig:THA_EOPM_sim_setup}
\end{figure*}

Reflectometry traces, such as the Luna \ac{OBR} traces, take on the order of a few seconds, which are not fast enough to measure, compute and gain information of the key during a \ac{QKD} exchange since the modulators of an on-chip transmitter usually operate above MHz speeds. 
In addition, Eve must learn both intensity and phase encoding settings to allow her to reconstruct the full key. Fast readout of phase information can be obtained using a homodyne detector instead. A schematic is depicted in figure \ref{fig:THA_EOPM_sim_setup}. This setup was used in the simulation software PICWAVE using chip component models provided for components of the Fraunhofer \ac{HHI} \ac{InP} \ac{PDK} \cite{PICWaveHHI2022}. 

\begin{figure}[h]
	\centering
	\includegraphics[width=0.95\linewidth]{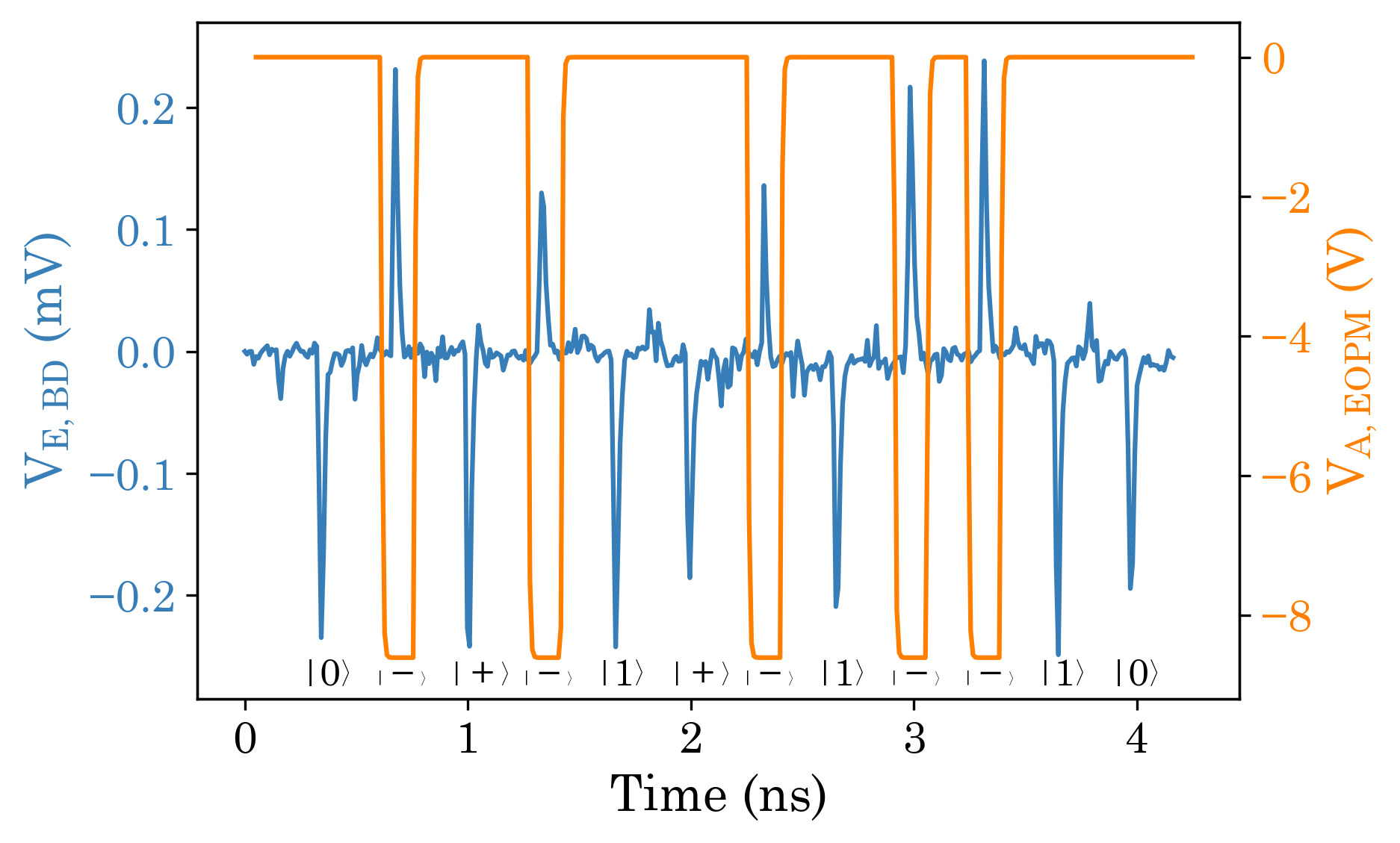}
	\caption[THA on HHI EOPM with PICWave simulation]{Successful \ac{THA} on an on-chip \ac{EOPM}. Alice encodes either 0 or $\pi$ phase change, depending on the state she sends, with a stand-alone \ac{EOPM} using voltage $V_{A,EOPM}$. Eve's back reflected light has a corresponding phase change of either 0 or $\pi$, which gives either a positive or a negative voltage peak at Eve's balanced detector, $V_{E,BD}$. Positive peaks are observed for $\pi$ phase change, i.e. a $\ket{-}$ state, negative voltage peaks are observed for 0 phase change, so $\ket{0}$, $\ket{1}$ and $\ket{+}$ states.
	}
	\label{fig:THA_EOPM_sim}
\end{figure}

The simulation set-up uses a simplified photonic circuit, consisting of only an \ac{EOPM} and two \acp{SSC}. The homodyne detector set-up uses an \ac{AMZI} to interfere the high reflection peak from the chip entrance at $\rm SSC_f$, serving as a local oscillator, with the light that has gone through the chip and reflected at $\rm SSC_b$, the signal\footnote{In an experiment one would use a circulator in place of $\rm BS_1$, however our software does not include such models since it is made for integrated photonics.}. A phase modulator, a \ac{TOPM} in the simulation, was used to tune the homodyne detector to have a maximum peak difference between different \ac{EOPM} settings. In phase encoding of time bin qubits, a phase of $\pi$ distinguishes the $\ket{+}$ and $\ket{-}$ states. Both states have the same light intensity in both time bins, but for $\ket{+}$ the pulses in both time bins have the same phase, whereas for $\ket{-}$ they are out of phase. If the phase modulator on chip was either permanently switched to 0 or $\pi$, the homodyne detector would not be able to distinguish between the states, since Eve's light would pass the modulator twice and would obtain either a $2\times0$ or a $2\times\pi$ phase. However, when operating the chip at high speeds, Eve can ensure her light only passes through the modulator once when there is a chance it is set to create a $\pi$ phase. The group index of an \ac{InP} waveguide is around 3.7. On chip-modulators are rated for bandwidths up to 40 GHz, so during one switching cycle, a pulse would propagate $\sim 2$ mm. This is on the order of chip size, so at high modulation speeds it is certainly realistic to assume Eve can ensure her light only passes the phase modulator once when switched to encode $\pi$ phase.

The simulation results are shown in figure \ref{fig:THA_EOPM_sim}, including both the electrical pulses Alice could use to encode a $\pi$ phase shift, and the measured voltage at Eve's homodyne detector. Alice's \ac{EOPM} has a $V_{\pi} = - 8.61$ V. For the states $\ket{0}$, $\ket{1}$ and $\ket{+}$ Alice does not encode a phase in either time bin. For the state $\ket{-}$, Alice switches the \ac{EOPM} voltage from 0 to $V_{\pi}$ between the two-time bins.

One can see that Eve's homodyne detector registers a negative voltage peak when Alice encodes no phase change between the two time bins and a positive peak when there is a phase change. Therefore, it is important to note that Eve can distinguish between the two settings of the phase modulator that Alice uses. 
    
\section{Securing QKD chips} \label{sec:securing}
Securing attacks on Alice's transmitter can be challenging, since the intensity of the light Eve uses varies from just a few photons, as for example in a \ac{THA} as described in the last section, to enough light to damage the chip permanently as in some \acp{LDA}. Since chip scale QKD systems have shown resilience against \acp{LDA} \cite{Ruzhitskaya2020,Joehlinger2022}, with the attack often being obvious from the precise temperature control needed for operating the chips, we will concentrate on the case of \ac{THA}. 
In the following we will describe hacking prevention mechanisms for Eve attacking Alice's transmitter chip, in the case of \ac{MDI} \ac{QKD} Eve's attacks and their prevention apply equivalently to Bob's transmitter system.

\subsection{Hacking prevention scheme}

\begin{figure}[h!]
	\centering
	\includegraphics[width=0.95\linewidth]{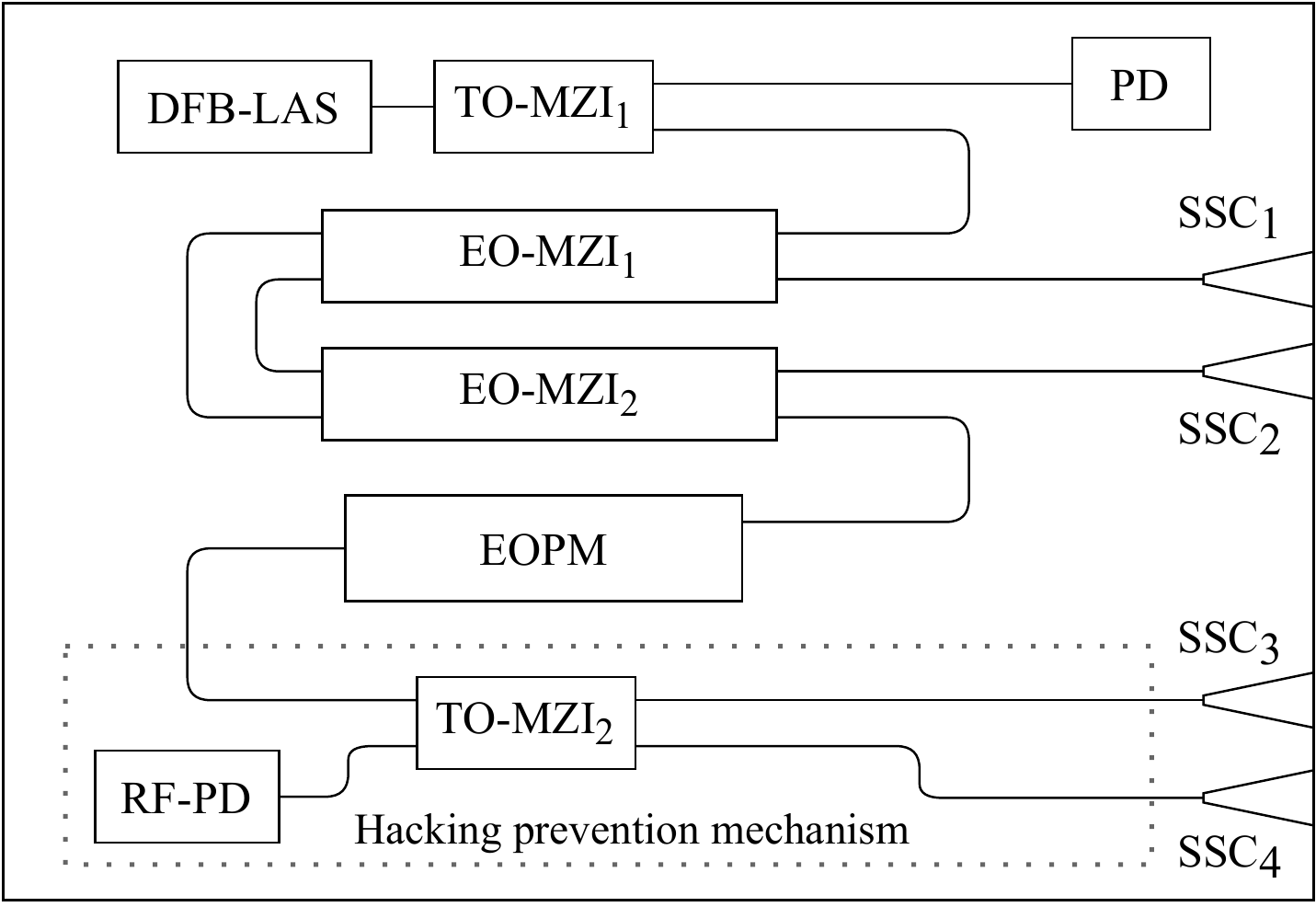}
	\caption[Layout of new MDI chip]{Optical circuit of the new \ac{MDI} \ac{QKD} chip design. The chip contains one main circuit consisting of a laser source, a \ac{DFB} laser, followed by an \ac{MZI}. One output path of this \ac{MZI} leads to a low speed photodiode (PD) for characterisation and monitoring purposes, the other output arm leads to two RF \acp{MZI} and a stand-alone \ac{EOPM}. At the bottom, the hacking prevention circuit can be found, consisting of one \ac{MZI} and an RF photodiode. Each \ac{MZI} consists of \acp{MMI} on each end and phase modulators in each arm: \acp{TOPM} for the TO-MZIs and \acp{EOPM} for the faster EO-MZIs.}
	\label{fig:HHI29-MDI}
\end{figure}

Let $I_{E,min,out}$ be the minimum amount of light that Eve needs to receive back to gain information about Alice's encoding and $I_{E,out}$ the amount she actually receives back. Figure \ref{fig:HHI29-MDI} shows an example \ac{MDI} \ac{QKD} transmitter chip with the hacking prevention circuit at the bottom. The goal of the hacking prevention mechanism is to ensure $I_{E,out} < I_{E,min,out}$, i.e. Eve receives less light back than she needs to gain significant amounts of information. Another thing to note is that Eve's light cannot gain intensity on reflection, so $I_{E,in} \geq I_{E,min,out}$. \\

If we capture the total on-chip reflectivity of the encoding system in $R_{enc}$ (excluding the fibre-to-chip coupling efficiency, $\epsilon_{oc}$, and any hacking prevention circuit), then in the case of an unprotected chip, if Eve sends in light of intensity $I_{E,in}$ she receives back

\begin{equation}
I_{E,out} = R_{enc} \times \epsilon_{oc}^2 \times I_{E,in} .
\end{equation}

If we also consider the ratio of the variable beamsplitter in the hacking defense mechanism $r_{enc} = 1 - r_{PD}$, where $r_{PD}$ is the ratio of light reaching the photodiode in the defense mechanism, then Eve receives back

\begin{equation}
    I_{E,out} = R_{enc} \times \epsilon_{oc}^2 \times I_{E,in} \times r_{enc}^2
\end{equation}

Let $I_{PD,min}$ describe the minimum light intensity the photodiode can detect. Alice can detect Eve if Eve's light reaching the photodiode is higher than $I_{PD,min}$, i.e. 

\begin{equation}
I_{E,in} \times r_{PD} \times \epsilon_{oc} > I_{PD,min} 
\end{equation}

Combining these equations and inequalities we can ensure that if the following inequality holds, Eve would not be able to gather sufficient information to allow her to fully reconstruct the secret key:

\begin{equation}
    \frac{r_{enc}}{r_{PD}} <
    \frac{I_{E,min,out}}
         {I_{PD,min}\sqrt{R_{enc}}}
\end{equation}

As an example consider an \ac{InP} chip made by the foundry \ac{HHI}. We can estimate the lowest detectable light intensity from the dark current, given as $i_{dark}< 10 $ nA by \ac{HHI} \cite{HHI2022}, and responsivity $\mathcal{R}$ of the photodiode, which was found to be $\mathcal{R} = 0.89$ A/W in simulation. This gives us $I_{PD,min} = i_{dark}/\mathcal{R} < 11 $ nW.

Eve can gain information from a single photon returning to her, which means $I_{E,min,out} = E_{\gamma} \times f_{rep}$. Using $ E_{\gamma} = 1.282\times10^{-19}$ J for light at 1550nm and a 7.7 GHz repetition rate $f_{rep}$, see next section, we obtain $I_{E,min,out} = 0.99$ nW.

Finally we can use the given reflectivity of the \acp{SSC} as the highest reflectivity expected in the chip, $R_{enc} = 0.001$, using the standard anti-reflectivity coating \cite{HHI2022}. Combining these three values we can see that the beamsplitter ratio must be set to 

\begin{equation}
    \frac{r_{enc}}{r_{PD}} < 2.85.
\end{equation}

This can be achieved with an \ac{MZI} variable beam splitter, such as shown in figure \ref{fig:HHI29-MDI}. At this beam splitting ratio, enough of Eve's light is directed towards the RF photodiode that, if Eve sends in enough light to allow her to reconstruct the key, she will be detected by Alice's photodiode.

\section{Benefit of High Speed} \label{sec:highspeed}

\begin{figure}[h]
    \centering
    \begin{subfigure}[t]{\linewidth}
        \centering
        \includegraphics[width=\linewidth]{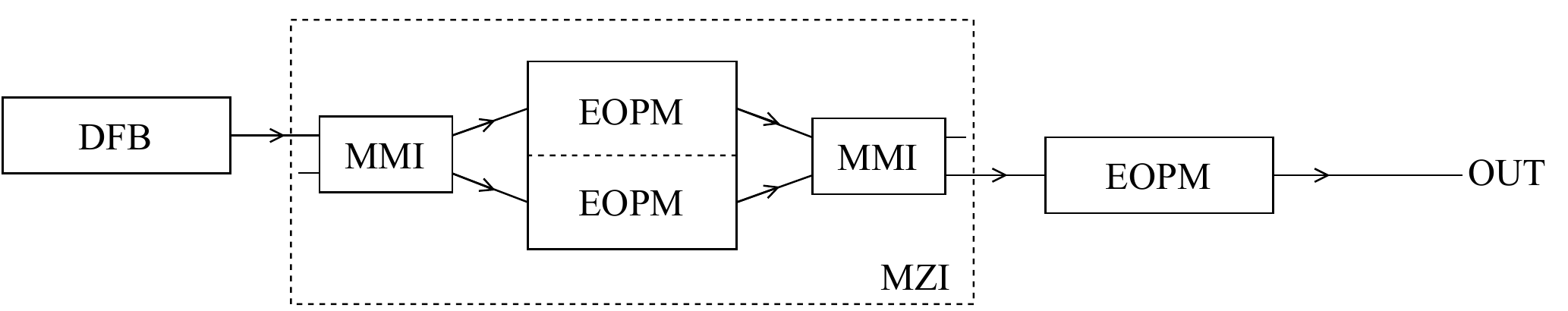}
        \caption{}
    \end{subfigure}
    \begin{subfigure}[t]{\linewidth}
        \includegraphics[width=\linewidth]{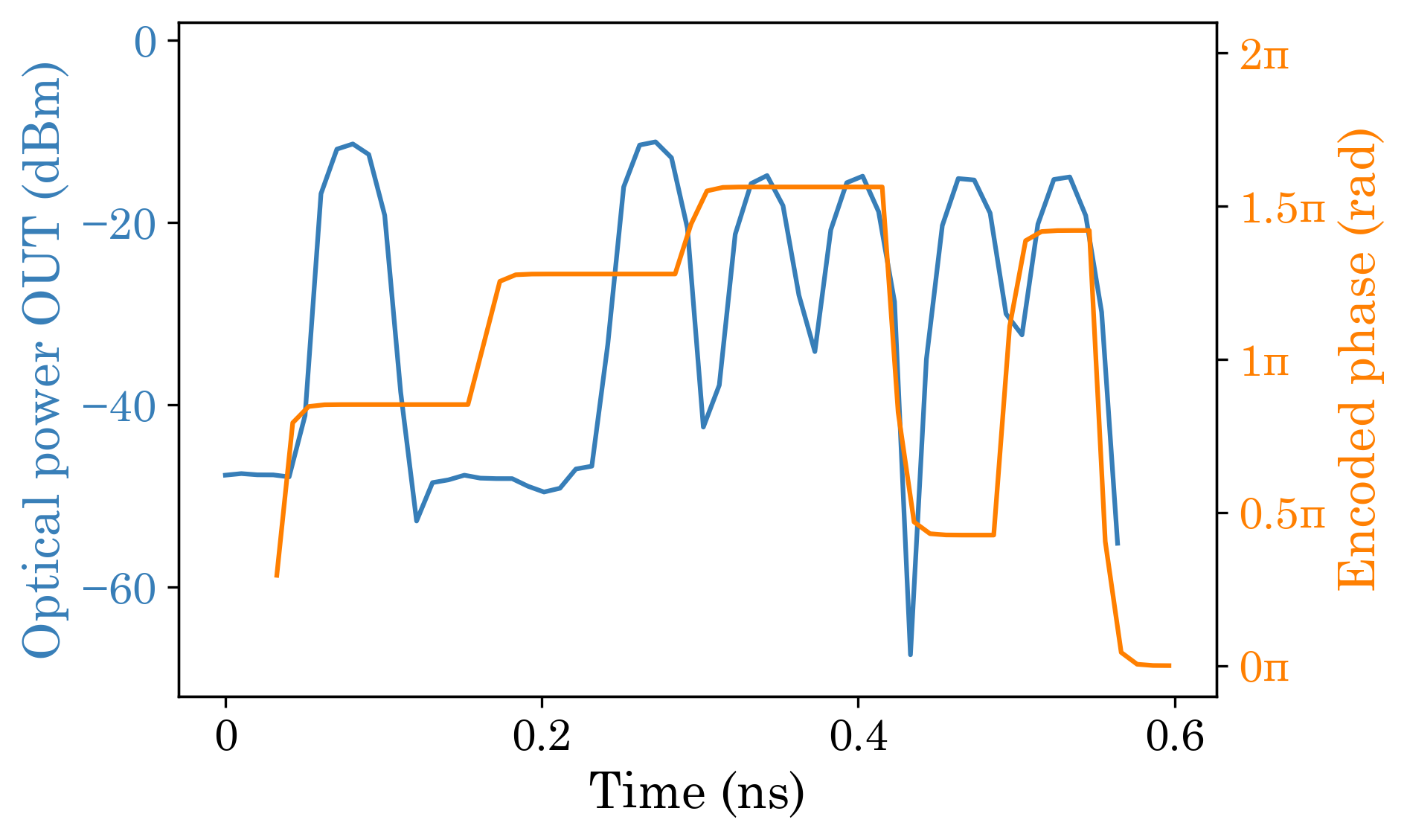}
        \caption{}
    \end{subfigure}
    \caption{\textbf{(a)} Simulation setup used for the BB84 \ac{QKD} state generation simulations in PICWave. The \ac{QKD} transmitter chip contains one \ac{DFB} laser, one \ac{MZI} and one stand-alone \ac{EOPM}. The \ac{MZI} consists of a twin \ac{EOPM} in between two \acp{MMI}. \textbf{(b)} Simulated examples of \ac{MDI} states created using a continuously current-driven \ac{DFB} laser showing the intensity of the created pulses together with the encoded phase, including both the discrete random phase and $\pi$ phase encoding, based on the voltage over the standalone \ac{EOPM}. Electrical signals were limited to a rise time of 4.5 ps. The sequence shown is $\ket{0}\ket{1}\ket{+}\ket{-}$.}
    \label{fig:MDI_sim_fast}
\end{figure}

Further simulations were performed in PICWAVE using the simplified setup shown in figure \ref{fig:MDI_sim_fast}(a) to understand the effect of speed on hacking prevention. Here, the \ac{DFB} laser was driven in the continuous mode operation instead of pulsed mode. In the pulsed mode operation, the laser pulse repetition rate is the transmitter speed, and also the laser is often driven using gain-switching to obtain the phase randomness required by most security proofs \cite{Lo2005}. However, this has a significant impact on the speed at which the chip can be operated, since one has to wait for the laser to fully turn off and turn back on again for true random phases. Instead, one can use active phase randomisation using a phase modulator. For \ac{MDI} \ac{QKD}, for example, one can approach the theoretical random phase secure key rate with only 14 discrete random phases \cite{Cao2020}. This is the approach of our simulation. 

\subsection{Simulation setup} \label{sec:highspeedsim}

It is common to separate the quantum state encoding into individual components. This simplifies the characterisation of the components and allows control using digital signals resulting in higher state fidelity. However, in simulation we are less restricted by this, so a simpler simulation setup with only a single \ac{MZI} and stand-alone \ac{EOPM} was used, shown in figure \ref{fig:MDI_sim_fast}(a). The phase encoding and discrete random phase modulation were performed by the \acp{EOPM}, whereas the \ac{MZI} was used for pulse carving and modulation. 

At high speed, we not only have to consider the speed at which the phase modulators can be operated, but also the speed of the control electronics. For example, the rise time of an \ac{AWG} is likely to be in the 10s of picoseconds. For the simulations we use the rise time of a state-of-the-art \ac{AWG}, the Keysight M8199A \ac{AWG}, of 4.5 ps. The results of these simulations are shown in figure \ref{fig:MDI_sim_fast}(b). The total time for each symbol was 130 ps: 30 ps for each pulse in the the symbol, with 30 ps in between the two pulses and 40 ps between pulses in different symbols. In previous implementations an extinction of 30 dB between symbols and 20 dB between pulses in the two time-bins in the same symbol were achieved \cite{Semenenko2020}. We can also see this in the simulation results here, indicating good performance at a symbol rate of $f_{rep} = 7.7$ GHz. If demonstrated, this would be an improvement of more than 6 times compared to previous \ac{MDI} \ac{QKD} implementation \cite{Wei2020} and an improvement of more than 30 times compared to previous on-chip time-bin encoded \ac{MDI} \ac{QKD} systems \cite{Semenenko2020}.

\subsection{Security from high speed} \label{sec:highspeedsecurity}

\begin{figure*}[t]
    \centering
    \begin{subfigure}[t]{0.45\linewidth}
        \centering
        \includegraphics[width=\linewidth]{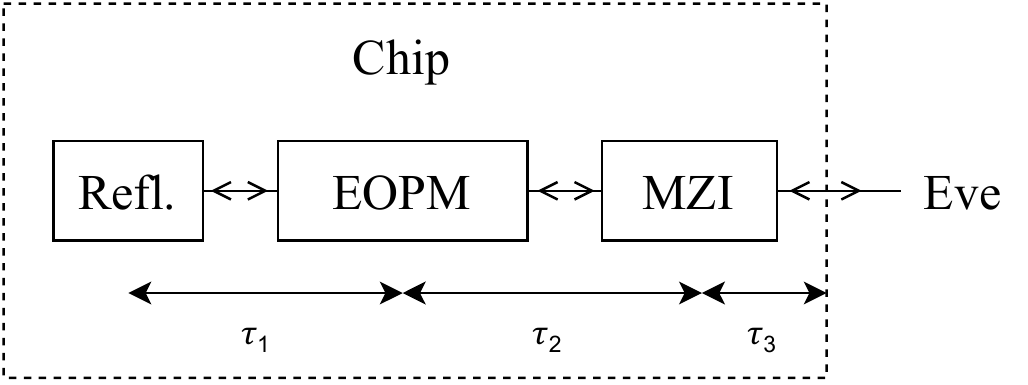}
        \caption{}
    \end{subfigure}
    \begin{subfigure}[t]{0.45\linewidth}
        \includegraphics[width=\linewidth]{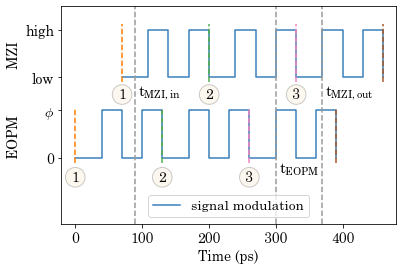}
        \caption{}
    \end{subfigure}
    \begin{subfigure}[t]{0.45\linewidth}
        \includegraphics[width=\linewidth]{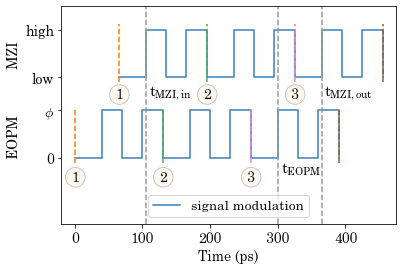}
        \caption{}
    \end{subfigure}
    \begin{subfigure}[t]{0.45\linewidth}
        \includegraphics[width=\linewidth]{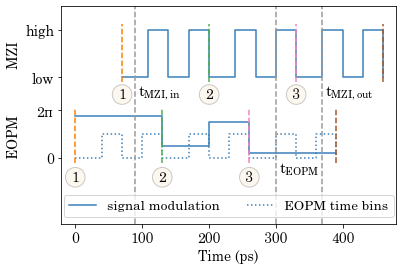}
        \caption{}
    \end{subfigure}
    \caption{
    Modulation speed and timing as a security measure: Three \ac{QKD} symbols marked 1,2,3, each containing two time bins, are shown passing through the components as seen in (a) with different optical paths (time delays) between the on-chip components. The set of figures shows that with certain back reflection delays (as in (b) and (d)), Eve's pulse will not be able to return any modulation information through the amplitude modulator's "on" time slots. \textbf{(a)} Example of an encoding component sequence on a chip with a single reflection point (Refl.) and the optical path lengths between them. \textbf{(b)} Time bins for \ac{EOPM} and \ac{MZI} for $\tau_1 = 70$ ps and $\tau_2 = 70$ ps. The modulation period is 30 ps with 40 ps between symbols. $t_{MZI,in}$ is the time Eve's light must encounter the \ac{MZI} for the first time, such that both Alice's pulse and Eve's reflected pulse go forward through the \ac{EOPM} at $t_{EOPM}$, and through the \ac{MZI} at $t_{MZI,out}$. At $t_{MZI,in}$, Alice's \ac{MZI} is "off", so Eve cannot enter the chip at the right time and the \ac{THA} fails. \textbf{(c)} Similar to (b), except $\tau_1$ and $\tau_2$ are 65 ps. In this version, Alice's \ac{MZI} is "on" at $t_{MZI,in}$, so Eve can perform a \ac{THA}. \textbf{(d)} Similar to (b), except for the switching signals for the \ac{EOPM} being adapted for high speed with significant rise times. The \ac{EOPM} encodes both random phase and $\pi$ phase change between time bins for $\ket{-}$ states. Despite having more time to measure the \ac{EOPM} state, Eve's \ac{THA} fails, since she still is restricted by the requirement to time her pulses such that they can pass the \ac{MZI} when it is "on" on both the inward and reflected path.
    }
    \label{fig:MDI_RF_refl_secure}
\end{figure*}

The \ac{THA} relies on measuring the state of the on-chip modulators using back reflections caused by other components further along the optical path. If the modulator closest to Eve is an on-off amplitude modulator, then any useful signal Eve sends must be timed such that both the transmitted and back-reflected pulses pass through the amplitude modulator during the time(s) it is "on". Provided that the amplitude modulator has sufficient extinction, and a fast enough rise time and repetition rate and that the optical path length to the back reflection(s) are chosen correctly, Eve will not be able to perform a \ac{THA}. 

Figure \ref{fig:MDI_RF_refl_secure}(a) shows a simple model with a source of back reflection, an \ac{EOPM} and an \ac{MZI} (acting like an amplitude modulator). The key information here is the time it takes for a photon or a classical light pulse to travel the distance from one modulation component to the next component within the chip. An \ac{EOPM} can be multiple milimeters long. Using a group index of about 3.7 for InP waveguides from foundries, we find that a pulse of light travels about 1 mm per 12 ps. 

For example, figure \ref{fig:MDI_RF_refl_secure}(a) shows the time intervals needed for light to go through different components of a chip placed at carefully calculated distances. Let's assume $\tau_1 = 70$ ps, $\tau_2 = 70$ ps and $\tau_3 = 40$ ps. In this example we assume a simple system, where the \ac{EOPM} and \ac{MZI} elements show no significant reflection points, but there is a significant point of reflection behind the \ac{EOPM}. This means that Eve's light will enter the chip, go through the \ac{MZI} and then the \ac{EOPM} after which it reflects back, encounters the \ac{EOPM} once more, followed by the \ac{MZI} before exiting the chip again. 

In order to reach the \ac{EOPM} and \ac{MZI} on the outwards path at the same time as Alice's light, Eve's light must enter the chip 320 ps before $t_{EOPM}$, the time Alice's light enters the \ac{EOPM}. This means she encounters the \ac{MZI} for the first time 210 ps before $t_{EOPM}$. The \ac{MZI} is set to minimum throughput in between pulses, since it is used for pulse carving. This means that if we have the time bin pattern shown in figure \ref{fig:MDI_RF_refl_secure}(b), the \ac{MZI} is at low throughput if Eve tries to inject her light to reach the \ac{EOPM} on reflection at the same time as Alice's light does and Eve's light never even makes it past the \ac{MZI}. Since Eve cannot change the speed of light, this means she cannot measure the \ac{EOPM} or \ac{MZI} settings hence the hacking prevention.

In contrast, figure \ref{fig:MDI_RF_refl_secure}(c) shows the placement of the components ($\tau_1 = 65 ps$, $\tau_2 = 65 ps$ and $\tau_3 = 40 ps$) when the \ac{MZI} is at high throughput at both $t_{MZI,in}$ and $t_{MZI,out}$ and thus Eve gets back the reflected light.

One important caveat in this analysis is that a chip rarely contains only one particular point of reflection, however, the same above method can be applied to more complex systems with careful design. A potentially more problematic drawback is the slow rise time of the modulators. At high speeds the low throughput settings of the \ac{MZI} may be quite short and with less extinction than the \ac{MZI} may be capable of, see for example the optical power graph in figure \ref{fig:MDI_sim_fast}(b), specifically the $\ket{+}$ and $\ket{-}$ states. In addition, to be able to accurately encode the $\pi$ phase changes between time bins in the $\ket{-}$ state and the random phases for all symbols, the standalone \ac{EOPM} is switched to the next phase as soon as the previous time bin is over, see figures \ref{fig:MDI_sim_fast}(b) and \ref{fig:MDI_RF_refl_secure}(d). Therefore, in principle, Eve has a wider time period to measure the phase in between time bins. To ensure proper extinction such that Eve receives no light fulfilling the condition set in section \ref{sec:securing} these restrictions from rise time must be taken into account and it may be safer to operate the chips at slightly lower speeds.

\section{Discussion} \label{sec:discussion}
Compared to other works on the security of chip-scale \ac{QKD} transmitters, our work presented above has some key differences. In \cite{Tan2021}, a secure \ac{MDI} \ac{QKD} implementation was created by including the information loss from the \ac{THA} in the secure key rate formula and using optical isolators to reduce the information Eve can obtain through a \ac{THA} and minimise this loss. This follows the general idea in the literature which is to reduce the information Eve obtains by setting a physical maximal intensity of light (12.8W) that Eve can inject without damaging the fibre and then adding optical isolation to reduce the light Eve receives back and use this to reduce the information Eve can obtain of the key. In \cite{Tan2021} multiple isolators in series (with a total of 232 dB isolation) also provides some protection against a \ac{LDA} attack, however, optical isolators may still be vulnerable to \acp{LDA} at other wavelengths \cite{Makarov2023}.
Using isolators as a defense is passive, preventing an attack at all times, however, with no record of whether there was an attack. Our defense is active, which requires continuous monitoring of the photodiode on chip, but also alerts us to an attack. The photodiode is technically vulnerable to an \ac{LDA}, however, so far chips have shown to break coupling before the photodiode is much affected, i.e. a \ac{DOS} attack \cite{Ruzhitskaya2020,Joehlinger2022}. 

Although \cite{Tan2021} concentrated on polarisation encoding \ac{MDI}, the results should still be applicable to \ac{MDI} in general. The authors used silicon chips with 2D grating couplers for coupling orthogonal polarisations, while we used \ac{InP} chips with \ac{SSC} couplers, however both are reflective from the perspective of an eavesdropper.  

In the above context, hacking prevention mechanisms of most chip-scale implementations will have similar elements: the maximum amount of light Eve can inject is bound by the coupling loss between fibre and chip, splitting ratio of the "defense beam splitter/MZI" and minimum light that the on-chip photodiode can detect. For a given chip with a known coupling loss and photodiode characteristics, the maximum amount of light Eve can inject is therefore subject to the value of the splitting ratio that has been set. Through setting this value, and knowing the maximum reflection in the chip, one should be able to restrict Eve’s information to prevent hacking.

\section{Conclusions} \label{sec:conclusions}

Integrated photonics is a promising option for commercialisation of \ac{QKD}, due to its low size, weight and power consumption, potential for high volume production at low cost and high interferometric stability. In this paper we have shown two options for securing on-chip \ac{QKD} transmitters against quantum hacking using only on-chip components. One can secure a transmitter chip using a combination of an \ac{MZI} and a photodiode against attacks such as the \ac{THA} to ensure Eve's attack is detected before she gains any significant amount of information. Additionally, we showed that due to the high speed at which chips can operate, and their small size and the precise placement of encoding components on the chip relative to each other, one can coordinate operating speeds and component placement on the chip to significantly reduce the information Eve receives. 
Combining both of these methods allows for a more secure \ac{QKD} transmitter, which, if combined with the \ac{MDI} \ac{QKD} protocol, creates a highly secure \ac{QKD} implementation. Here we showed these methods in principle, future work is needed to include them in a secure key rate formula to allow them to be used to secure real-world \ac{QKD} systems.
\section*{Acknowledgements} \label{sec:acknowledgements}
    The authors acknowledge financial support from Innovate UK project 10102791 QAssure, Engineering and Physical Science Research Council (EPSRC) Quantum Communications Hubs EP/M013472/1 and EP/T001011/1, EPSRC training grant EP/LO15730/1, and EPSRC EP/N015126/1 Quantum Photonic Integrated Circuits (QuPIC).
\section*{Author Contributions}
The experiments were conducted by FJ with help from HS and planning and supervision from CE, DA, JR and PS. Simulations were carried out by FJ under the supervision of SJ, IF and JR. Key concepts of the shown hacking prevention mechanisms were  developed by FJ, supervised by IF, SJ and JR. The paper was written by FJ with help from IF and SJ. CE, JR and SJ contributed to the grants that supported this work. All authors discussed the results and commented on the manuscript.
\FloatBarrier



\begin{thebibliography}{10}
\expandafter\ifx\csname url\endcsname\relax
  \def\url#1{\texttt{#1}}\fi
\expandafter\ifx\csname urlprefix\endcsname\relax\def\urlprefix{URL }\fi
\providecommand{\bibinfo}[2]{#2}
\providecommand{\eprint}[2][]{\url{#2}}

\bibitem{BB84}
\bibinfo{author}{Bennett, C.} \& \bibinfo{author}{Brassard, G.}
\newblock \bibinfo{title}{Quantum cryptography: Public key distribution and
  coin tossing}.
\newblock \emph{\bibinfo{journal}{Proceedings of the IEEE International
  Conference on Computers, Systems and Signal Processing}}
  \bibinfo{pages}{175--179} (\bibinfo{year}{1984}).

\bibitem{Lo1999}
\bibinfo{author}{Lo, H.-K.} \& \bibinfo{author}{Chau, H.~F.}
\newblock \bibinfo{title}{Unconditional security of quantum key distribution
  over arbitrarily long distances}.
\newblock \emph{\bibinfo{journal}{Science}} \textbf{\bibinfo{volume}{283}},
  \bibinfo{pages}{2050–2056} (\bibinfo{year}{1999}).
\newblock \urlprefix\url{http://dx.doi.org/10.1126/science.283.5410.2050}.

\bibitem{Braunstein2012}
\bibinfo{author}{Braunstein, S.~L.} \& \bibinfo{author}{Pirandola, S.}
\newblock \bibinfo{title}{Side-channel-free quantum key distribution}.
\newblock \emph{\bibinfo{journal}{Phys. Rev. Lett.}}
  \textbf{\bibinfo{volume}{108}}, \bibinfo{pages}{130502}
  (\bibinfo{year}{2012}).
\newblock
  \urlprefix\url{https://link.aps.org/doi/10.1103/PhysRevLett.108.130502}.

\bibitem{Lo2012}
\bibinfo{author}{Lo, H.-K.}, \bibinfo{author}{Curty, M.} \&
  \bibinfo{author}{Qi, B.}
\newblock \bibinfo{title}{Measurement-device-independent quantum key
  distribution}.
\newblock \emph{\bibinfo{journal}{Physical Review Letters}}
  \textbf{\bibinfo{volume}{108}} (\bibinfo{year}{2012}).
\newblock \urlprefix\url{http://dx.doi.org/10.1103/PhysRevLett.108.130503}.

\bibitem{Makarov2006}
\bibinfo{author}{Makarov, V.}, \bibinfo{author}{Anisimov, A.} \&
  \bibinfo{author}{Skaar, J.}
\newblock \bibinfo{title}{Effects of detector efficiency mismatch on security
  of quantum cryptosystems}.
\newblock \emph{\bibinfo{journal}{Physical Review A}}
  \textbf{\bibinfo{volume}{74}} (\bibinfo{year}{2006}).
\newblock \urlprefix\url{http://dx.doi.org/10.1103/PhysRevA.74.022313}.

\bibitem{Rogers2007}
\bibinfo{author}{Rogers, D.~J.}, \bibinfo{author}{Bienfang, J.~C.},
  \bibinfo{author}{Nakassis, A.}, \bibinfo{author}{Xu, H.} \&
  \bibinfo{author}{Clark, C.~W.}
\newblock \bibinfo{title}{Detector dead-time effects and paralyzability in
  high-speed quantum key distribution}.
\newblock \emph{\bibinfo{journal}{New Journal of Physics}}
  \textbf{\bibinfo{volume}{9}}, \bibinfo{pages}{319–319}
  (\bibinfo{year}{2007}).
\newblock \urlprefix\url{http://dx.doi.org/10.1088/1367-2630/9/9/319}.

\bibitem{Wiechers2011}
\bibinfo{author}{Wiechers, C.} \emph{et~al.}
\newblock \bibinfo{title}{After-gate attack on a quantum cryptosystem}.
\newblock \emph{\bibinfo{journal}{New Journal of Physics}}
  \textbf{\bibinfo{volume}{13}}, \bibinfo{pages}{013043}
  (\bibinfo{year}{2011}).
\newblock \urlprefix\url{http://dx.doi.org/10.1088/1367-2630/13/1/013043}.

\bibitem{Sauge2011}
\bibinfo{author}{Sauge, S.}, \bibinfo{author}{Lydersen, L.},
  \bibinfo{author}{Anisimov, A.}, \bibinfo{author}{Skaar, J.} \&
  \bibinfo{author}{Makarov, V.}
\newblock \bibinfo{title}{Controlling an actively-quenched single photon
  detector with bright light}.
\newblock \emph{\bibinfo{journal}{Optics Express}}
  \textbf{\bibinfo{volume}{19}}, \bibinfo{pages}{23590} (\bibinfo{year}{2011}).
\newblock \urlprefix\url{http://dx.doi.org/10.1364/OE.19.023590}.

\bibitem{Korzh2015}
\bibinfo{author}{Korzh, B.} \emph{et~al.}
\newblock \bibinfo{title}{Provably secure and practical quantum key
  distribution over 307 km of optical fibre}.
\newblock \emph{\bibinfo{journal}{Nature Photonics}}
  \textbf{\bibinfo{volume}{9}}, \bibinfo{pages}{163–168}
  (\bibinfo{year}{2015}).
\newblock \urlprefix\url{http://dx.doi.org/10.1038/nphoton.2014.327}.

\bibitem{Ursin2007}
\bibinfo{author}{Ursin, R.} \emph{et~al.}
\newblock \bibinfo{title}{Entanglement-based quantum communication over 144
  km}.
\newblock \emph{\bibinfo{journal}{Nature Physics}}
  \textbf{\bibinfo{volume}{3}}, \bibinfo{pages}{481–486}
  (\bibinfo{year}{2007}).
\newblock \urlprefix\url{http://dx.doi.org/10.1038/nphys629}.

\bibitem{Sibson2017b}
\bibinfo{author}{Sibson, P.} \emph{et~al.}
\newblock \bibinfo{title}{Chip-based quantum key distribution}.
\newblock \emph{\bibinfo{journal}{Nature Communications}}
  \textbf{\bibinfo{volume}{8}} (\bibinfo{year}{2017}).
\newblock \urlprefix\url{http://dx.doi.org/10.1038/ncomms13984}.

\bibitem{Paraiso2021}
\bibinfo{author}{Paraïso, T.~K.} \emph{et~al.}
\newblock \bibinfo{title}{A photonic integrated quantum secure communication
  system}.
\newblock \emph{\bibinfo{journal}{Nature Photonics}}
  \textbf{\bibinfo{volume}{15}}, \bibinfo{pages}{850–856}
  (\bibinfo{year}{2021}).
\newblock \urlprefix\url{http://dx.doi.org/10.1038/s41566-021-00873-0}.

\bibitem{Scarani2014}
\bibinfo{author}{Scarani, V.} \& \bibinfo{author}{Kurtsiefer, C.}
\newblock \bibinfo{title}{The black paper of quantum cryptography: Real
  implementation problems}.
\newblock \emph{\bibinfo{journal}{Theoretical Computer Science}}
  \textbf{\bibinfo{volume}{560}}, \bibinfo{pages}{27--32}
  (\bibinfo{year}{2014}).
\newblock
  \urlprefix\url{https://www.sciencedirect.com/science/article/pii/S0304397514006938}.
\newblock \bibinfo{note}{Theoretical Aspects of Quantum Cryptography –
  celebrating 30 years of BB84}.

\bibitem{Sun2015}
\bibinfo{author}{Sun, S.-H.} \emph{et~al.}
\newblock \bibinfo{title}{Effect of source tampering in the security of quantum
  cryptography}.
\newblock \emph{\bibinfo{journal}{Physical Review A}}
  \textbf{\bibinfo{volume}{92}} (\bibinfo{year}{2015}).
\newblock \urlprefix\url{http://dx.doi.org/10.1103/PhysRevA.92.022304}.

\bibitem{Jain2015}
\bibinfo{author}{Jain, N.} \emph{et~al.}
\newblock \bibinfo{title}{Risk analysis of trojan-horse attacks on practical
  quantum key distribution systems}.
\newblock \emph{\bibinfo{journal}{IEEE Journal of Selected Topics in Quantum
  Electronics}} \textbf{\bibinfo{volume}{21}}, \bibinfo{pages}{168–177}
  (\bibinfo{year}{2015}).
\newblock \urlprefix\url{http://dx.doi.org/10.1109/JSTQE.2014.2365585}.

\bibitem{Vakhitov2001}
\bibinfo{author}{Vakhitov, A.}, \bibinfo{author}{Makarov, V.} \&
  \bibinfo{author}{Hjelme, D.~R.}
\newblock \bibinfo{title}{Large pulse attack as a method of conventional
  optical eavesdropping in quantum cryptography}.
\newblock \emph{\bibinfo{journal}{Journal of Modern Optics}}
  \textbf{\bibinfo{volume}{48}}, \bibinfo{pages}{2023–2038}
  (\bibinfo{year}{2001}).
\newblock \urlprefix\url{http://dx.doi.org/10.1080/09500340108240904}.

\bibitem{Lowndes2021}
\bibinfo{author}{Lowndes, D.}, \bibinfo{author}{Frick, S.},
  \bibinfo{author}{Hart, A.} \& \bibinfo{author}{Rarity, J.}
\newblock \bibinfo{title}{A low cost, short range quantum key distribution
  system}.
\newblock \emph{\bibinfo{journal}{EPJ Quantum Technology}}
  \textbf{\bibinfo{volume}{8}} (\bibinfo{year}{2021}).
\newblock \urlprefix\url{http://dx.doi.org/10.1140/epjqt/s40507-021-00101-2}.

\bibitem{Sibson2017}
\bibinfo{author}{Sibson, P.} \emph{et~al.}
\newblock \bibinfo{title}{Integrated silicon photonics for high-speed quantum
  key distribution}.
\newblock \emph{\bibinfo{journal}{Optica}} \textbf{\bibinfo{volume}{4}},
  \bibinfo{pages}{172} (\bibinfo{year}{2017}).
\newblock \urlprefix\url{http://dx.doi.org/10.1364/OPTICA.4.000172}.

\bibitem{Bunandar2018}
\bibinfo{author}{Bunandar, D.} \emph{et~al.}
\newblock \bibinfo{title}{Metropolitan quantum key distribution with silicon
  photonics}.
\newblock \emph{\bibinfo{journal}{Phys. Rev. X}} \textbf{\bibinfo{volume}{8}},
  \bibinfo{pages}{021009} (\bibinfo{year}{2018}).
\newblock \urlprefix\url{https://link.aps.org/doi/10.1103/PhysRevX.8.021009}.

\bibitem{Sajeed2021}
\bibinfo{author}{Sajeed, S.} \emph{et~al.}
\newblock \bibinfo{title}{An approach for security evaluation and certification
  of a complete quantum communication system}.
\newblock \emph{\bibinfo{journal}{Scientific Reports}}
  \textbf{\bibinfo{volume}{11}}, \bibinfo{pages}{5110} (\bibinfo{year}{2021}).
\newblock \urlprefix\url{https://doi.org/10.1038/s41598-021-84139-3}.

\bibitem{idQ}
\bibinfo{title}{What are the cybersecurity risks to current cryptographic
  techniques?}
\newblock \bibinfo{type}{Tech. Rep.}, \bibinfo{institution}{idQuantique}.
\newblock
  \urlprefix\url{https://www.idquantique.com/quantum-safe-security/quantum-key-distribution/}.

\bibitem{Lucamarini2015}
\bibinfo{author}{Lucamarini, M.} \emph{et~al.}
\newblock \bibinfo{title}{Practical security bounds against the trojan-horse
  attack in quantum key distribution}.
\newblock \emph{\bibinfo{journal}{Phys. Rev. X}} \textbf{\bibinfo{volume}{5}},
  \bibinfo{pages}{031030} (\bibinfo{year}{2015}).
\newblock \urlprefix\url{https://link.aps.org/doi/10.1103/PhysRevX.5.031030}.

\bibitem{Bugge2014}
\bibinfo{author}{Bugge, A.~N.} \emph{et~al.}
\newblock \bibinfo{title}{Laser damage helps the eavesdropper in quantum
  cryptography}.
\newblock \emph{\bibinfo{journal}{Physical Review Letters}}
  \textbf{\bibinfo{volume}{112}} (\bibinfo{year}{2014}).
\newblock \urlprefix\url{http://dx.doi.org/10.1103/PhysRevLett.112.070503}.

\bibitem{Jain2014}
\bibinfo{author}{Jain, N.} \emph{et~al.}
\newblock \bibinfo{title}{Trojan-horse attacks threaten the security of
  practical quantum cryptography}.
\newblock \emph{\bibinfo{journal}{New Journal of Physics}}
  \textbf{\bibinfo{volume}{16}}, \bibinfo{pages}{123030}
  (\bibinfo{year}{2014}).
\newblock \urlprefix\url{http://dx.doi.org/10.1088/1367-2630/16/12/123030}.

\bibitem{Sajeed2017}
\bibinfo{author}{Sajeed, S.}, \bibinfo{author}{Minshull, C.},
  \bibinfo{author}{Jain, N.} \& \bibinfo{author}{Makarov, V.}
\newblock \bibinfo{title}{Invisible trojan-horse attack}.
\newblock \emph{\bibinfo{journal}{Scientific Reports}}
  \textbf{\bibinfo{volume}{7}} (\bibinfo{year}{2017}).
\newblock \urlprefix\url{http://dx.doi.org/10.1038/s41598-017-08279-1}.

\bibitem{Tan2021}
\bibinfo{author}{Tan, H.}, \bibinfo{author}{Li, W.}, \bibinfo{author}{Zhang,
  L.}, \bibinfo{author}{Wei, K.} \& \bibinfo{author}{Xu, F.}
\newblock \bibinfo{title}{Chip-based quantum key distribution against
  trojan-horse attack}.
\newblock \emph{\bibinfo{journal}{Phys. Rev. Appl.}}
  \textbf{\bibinfo{volume}{15}}, \bibinfo{pages}{064038}
  (\bibinfo{year}{2021}).
\newblock
  \urlprefix\url{https://link.aps.org/doi/10.1103/PhysRevApplied.15.064038}.

\bibitem{Makarov2023}
\bibinfo{author}{Makarov, V.} \emph{et~al.}
\newblock \bibinfo{title}{Preparing a commercial quantum key distribution
  system for certification against implementation loopholes}
  (\bibinfo{year}{2023}).
\newblock \eprint{2310.20107}.

\bibitem{Ponosova2022}
\bibinfo{author}{Ponosova, A.} \emph{et~al.}
\newblock \bibinfo{title}{Protecting fiber-optic quantum key distribution
  sources against light-injection attacks}.
\newblock \emph{\bibinfo{journal}{PRX Quantum}} \textbf{\bibinfo{volume}{3}}
  (\bibinfo{year}{2022}).
\newblock \urlprefix\url{http://dx.doi.org/10.1103/PRXQuantum.3.040307}.

\bibitem{Ruzhitskaya2021}
\bibinfo{author}{Ruzhitskaya, D.} \emph{et~al.}
\newblock \bibinfo{title}{Vulnerabilities in the quantum key distribution
  system induced under a pulsed laser attack}.
\newblock \emph{\bibinfo{journal}{Scientific and Technical Journal of
  Information Technologies, Mechanics and Optics}}
  \textbf{\bibinfo{volume}{21}}, \bibinfo{pages}{837–847}
  (\bibinfo{year}{2021}).
\newblock
  \urlprefix\url{http://dx.doi.org/10.17586/2226-1494-2021-21-6-837-847}.

\bibitem{Semenenko2020}
\bibinfo{author}{Semenenko, H.} \emph{et~al.}
\newblock \bibinfo{title}{Chip-based measurement-device-independent quantum key
  distribution}.
\newblock \emph{\bibinfo{journal}{Optica}} \textbf{\bibinfo{volume}{7}},
  \bibinfo{pages}{238} (\bibinfo{year}{2020}).
\newblock \urlprefix\url{http://dx.doi.org/10.1364/optica.379679}.

\bibitem{PICWaveHHI2022}
\bibinfo{author}{{Photon Design}}.
\newblock \bibinfo{title}{{PICWave HHI PDK Instance - User Notes (v6.8.1.0)}}
  (\bibinfo{year}{2022}).

\bibitem{Ruzhitskaya2020}
\bibinfo{author}{Ruzhitskaya, D.} \emph{et~al.}
\newblock \bibinfo{title}{{Protecting QKD sources against light-injection
  attacks}} (\bibinfo{year}{2020}).
\newblock
  \urlprefix\url{https://2020.qcrypt.net/posters/QCrypt2020Poster043Huang.pdf}.

\bibitem{Joehlinger2022}
\bibinfo{author}{J{\"{o}}hlinger, F.}
\newblock \emph{\bibinfo{title}{{Security of Chip-Scale Quantum Key
  Distribution}}}.
\newblock Ph.D. thesis, \bibinfo{school}{University of Bristol}
  (\bibinfo{year}{2022}).

\bibitem{HHI2022}
\bibinfo{author}{{Fraunhofer HHI}}.
\newblock \bibinfo{title}{{HHI29 Design Manual (Version 6.9.0)}}
  (\bibinfo{year}{2022}).

\bibitem{Lo2005}
\bibinfo{author}{Lo, H.-K.} \& \bibinfo{author}{Preskill, J.}
\newblock \bibinfo{title}{Phase randomization improves the security of quantum
  key distribution} (\bibinfo{year}{2005}).
\newblock \urlprefix\url{https://arxiv.org/abs/quant-ph/0504209}.
\newblock \eprint{quant-ph/0504209}.

\bibitem{Cao2020}
\bibinfo{author}{Cao, Z.}
\newblock \bibinfo{title}{Discrete-phase-randomized
  measurement-device-independent quantum key distribution}.
\newblock \emph{\bibinfo{journal}{Physical Review A}}
  \textbf{\bibinfo{volume}{101}} (\bibinfo{year}{2020}).
\newblock \urlprefix\url{http://dx.doi.org/10.1103/PhysRevA.101.062325}.

\bibitem{Wei2020}
\bibinfo{author}{Wei, K.} \emph{et~al.}
\newblock \bibinfo{title}{High-speed measurement-device-independent quantum key
  distribution with integrated silicon photonics}.
\newblock \emph{\bibinfo{journal}{Physical Review X}}
  \textbf{\bibinfo{volume}{10}} (\bibinfo{year}{2020}).
\newblock \urlprefix\url{http://dx.doi.org/10.1103/physrevx.10.031030}.

\end{thebibliography}
\end{document}